\documentclass[sn-nature]{sn-jnl}% Style for submissions to Nature Portfolio journals
\usepackage{graphicx}%
\usepackage{multirow}%
\usepackage{amsmath,amssymb,amsfonts}%
\usepackage{amsthm}%
\usepackage{mathrsfs}%
\usepackage[title]{appendix}%
\usepackage{xcolor}%
\usepackage{textcomp}%
\usepackage{manyfoot}%
\usepackage{booktabs}%
\usepackage{algorithm}%
\usepackage{algorithmicx}%
\usepackage{algpseudocode}%
\usepackage{listings}%
\theoremstyle{thmstyleone}%
\theoremstyle{thmstyletwo}%

\theoremstyle{thmstylethree}%

\begin{document}

\title[STM-Net]{Reconstructing Pristine Molecular Orbitals from Scanning
Tunneling Microscopy Images via Artificial Intelligence Approaches}


\author[1]{\fnm{Yu} \sur{Zhu}}
\equalcont{These authors contributed equally to this work.}

\author[2]{\fnm{Renjie} \sur{Xue}}
\equalcont{These authors contributed equally to this work.}

\author[3]{\fnm{Hao} \sur{Ren}}

\author[1]{\fnm{Yicheng} \sur{Chen}}
\author[1]{\fnm{Wenjie} \sur{Yan}}
\author[1]{\fnm{Bingzheng} \sur{Wu}}

\author*[1,4]{\fnm{Sai} \sur{Duan}}\email{duansai@fudan.edu.cn}

\author*[2]{\fnm{Haiming} \sur{Zhang}}\email{hmzhang@suda.edu.cn}

\author*[2]{\fnm{Lifeng} \sur{Chi}}\email{chilf@suda.edu.cn}

\author*[1,4]{\fnm{Xin} \sur{Xu}}\email{xxchem@fudan.edu.cn}

\affil[1]{\orgdiv{Collaborative Innovation Center of Chemistry for
Energy Materials, Shanghai Key Laboratory of Molecular Catalysis
and Innovative Materials, MOE Key Laboratory of Computational Physical
Sciences, Department of Chemistry}, \orgname{Fudan University},
\orgaddress{\city{Shanghai}, \postcode{200433}, \country{P. R. China}}}

\affil[2]{\orgdiv{Jiangsu Key Laboratory for Carbon-Based Functional
Materials and Devices}, \orgname{Institute of Functional Nano and Soft
Materials (FUNSOM), Soochow University}, \orgaddress{\city{Suzhou},
\postcode{215123}, \state{Jiangsu}, \country{P. R. China}}}

\affil[3]{\orgdiv{School of Materials Science and Engineering},
\orgname{China University of Petroleum (East China)},
\orgaddress{\city{Qingdao}, \postcode{266580}, \state{Shandong},
\country{P. R. China}}}

\affil[4]{\orgname{Hefei National Laboratory}, \orgaddress{\city{Hefei},
\postcode{230088}, \country{P. R. China}}}


\abstract{Molecular orbital (MO) is one of the most fundamental concepts
for molecules, relating to all branches of chemistry, while scanning
tunneling microscopy (STM) has been widely recognized for its potential
to measure the spatial distribution of MOs. However, the
precise characterization of MO with high resolution in real space is a
long-standing challenge owing to the inevitable interference of
high-angular-momentum contributions from functionalized tips in STM.
Here, leveraging advances in artificial
intelligence for image recognition, we establish a physics-driven
deep-learning network, named STM-Net, to reconstruct MOs from
high-resolution STM images with a functionalized tip, taking advantage
of the separable characteristics of different angular momentum
contributions. We demonstrate that STM-Net can be directly applied to a
variety of experimental observations, successfully reconstructing
pristine MO features for molecules under diverse conditions.
Moreover, STM-Net can adapt to various states of the functionalized tip
and the substrate, illustrating the broad applicability of our
physics-driven framework. These results pave the way for accurate
characterization of MO with high resolution, potentially leading to new
insights and applications for this fundamental concept in chemistry.}

\maketitle

%\section{Introduction}\label{sec1}

Molecules are systems where nuclei are interconnected through chemical
bonds formed by shared electrons. These electrons act as a ``glue'',
counterbalancing the electrostatic repulsion between positively charged
nuclei and giving rise to molecular orbitals (MOs). As a fundamental
concept rooted in quantum mechanics\cite{mulliken1932electronic}, MOs
represent one of the most pivotal concepts in chemical science.
Specifically, frontier MOs, the highest occupied molecular orbital (HOMO)
and the lowest unoccupied molecular orbital (LUMO), play a critical role
in governing chemical reactivity in both ground and excited
states\cite{fukui1952molecular,woodward1965jacs}. Understanding the
distribution and energy levels of MOs provides a foundation for
addressing a wide range of chemical challenges. Despite ongoing debates
regarding the direct measurement of MOs\cite{mulder2011orbitals,pham2017can},
significant efforts\cite{zuo1999direct,repp2005molecules,gross2011recent}
have been devoted to developing experimental and theoretical approaches
to probe their characteristics by correlating observable phenomena with
MO characteristics\cite{kitou2017successive,tersoff1985theory}.

Among these techniques, scanning tunneling microscopy (STM) stands out
for its ability to provide high-resolution spatial information on MOs in
real space\cite{kitou2017successive,gross2011recent,repp2005molecules,chen2021introduction,duan2015jacs,duan2020jacs}.
By utilizing the energy conservation law during elastic electron
tunneling, STM can measure the resonance energies that correspond to the
energy levels of MOs. Furthermore, based on the Tersoff-Hamann
theory\cite{tersoff1985theory}, for sufficiently decoupled molecules,
constant-height imaging under an \textit{s}-wave tip (typically a
metallic tip) can be related to the MO's spatial distribution, yielding a
mapping $I_s(x,y)\propto|\phi(x,y,z_0)|^2$, where $I_s$ refers to the
\textit{s}-wave tunneling current, $\phi$ represents the MO at the
resonant energy level, and $\{x,y,z_0\}$ denotes the Cartesian
coordinate of the metallic tip apex\cite{tersoff1985theory,gross2011high,chen2021introduction}.
However, the inherent diffuse nature of the \textit{s}-wave, controlled
by the material's work function, imposes limitations on the resolution
achievable with metallic tips. While these tips produce physically
interpretable STM images\cite{tersoff1985theory}, their limited
resolution\cite{gross2011high} hinders a detailed, high-resolution
imaging of MOs.

Functionalized tips, on the other hand, offer enhanced resolution in
STM imaging, allowing for the visualization of MOs with finer
features\cite{gross2011high}. However, this enhancement comes at a cost,
such that high-angular-momentum contributions are inevitably introduced
to the tunneling current\cite{gross2011high,chen1990tunneling}. For
instance, the commonly used CO-decorated tip introduces a 
\textit{p}-wave contribution, which can be expressed as 
$I_p(x,y)\propto|\frac{\partial\phi(x,y,z_0)^2}{\partial{x}}|
+|\frac{\partial\phi(x,y,z_0)^2}{\partial{y}}|$, leading to undesirable
details other than MOs in STM images\cite{chen2021introduction,gross2011high,chen1990tunneling}.
In most cases, the contributions of $I_s$ (from the \textit{s}-wave) and
$I_p$ (from the \textit{p}-wave) to the total tunneling current of
CO-decorated tips are comparable\cite{gross2011high,duan2022general,duan2023accurate}.
Consequently, while functionalized tips enable high-resolution imaging,
the resultant images cannot be directly interpreted as the MO spatial
distribution. This creates a fundamental dilemma between achieving
high-resolution features with a functionalized tip and obtaining a
direct visualization of MOs with a low-resolution metallic tip.

Essentially, the aforementioned dilemma represents an image
transformation issue, i.e., how to disentangle and extract physically
meaningful components from a high-resolution image that contains
extraneous details. We note the unprecedented success of artificial
intelligence (AI) in image processing\cite{beyan2023review,castiglioni2021ai},
exemplified by breakthroughs such as AlexNet\cite{krizhevsky2012nips} in
the ImageNet challenge, which catalyzed the rapid advancement of deep
learning. While AI algorithms have achieved significant progress across
various domains\cite{yang2024mattersim,alberts2023large}, their most
transformative impact has been in the field of image processing.
Currently, advanced algorithms, such as U-Net\cite{ronneberger2015u},
generative adversarial networks\cite{goodfellow2020generative}, and
diffusion models\cite{ho2020denoising}, are widely applied to enhance
image resolution and quality. In the context of STM, however, the
application of AI algorithms has predominantly focused on optimizing
experimental workflows\cite{krull2020artificial,su2024natsyn,zhu2024jacs},
with limited exploration of its potential for interpreting STM images. A
major obstacle to this application is the interdependence of two
critical factors, namely, the lack of sufficient high-quality data and
the absence of tailored algorithms. In this work, we address this
challenge by developing STM-Net, a physics-driven deep learning network
that leverages advanced image segmentation algorithms to reconstruct pristine MO
distribution from high-resolution STM images obtained using
functionalized CO tips. This framework demonstrates exceptional
versatility, effectively handling complex molecular structures and
diverse experimental conditions.

\section*{Results and discussion}
\label{sec_rd}
\subsection*{Development of STM-Net}
\label{subsec_dstmnet}

\begin{figure}
\centering
\includegraphics[scale=0.75]{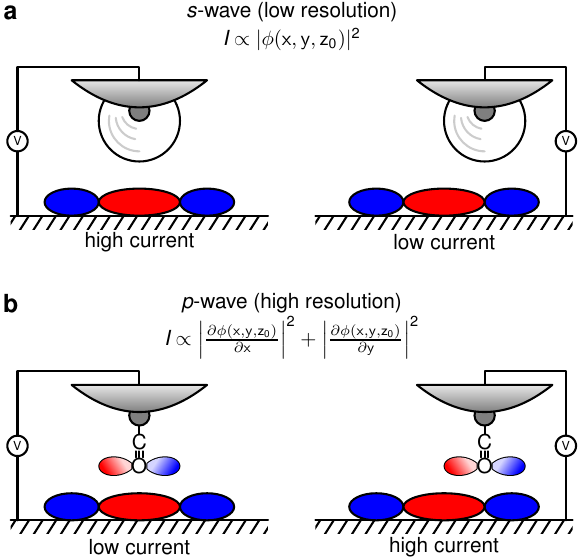}
\caption{\textbf{Characteristics of different angular-momentum
contributions}. \textbf{(a)} Illustration of an \textit{s}-wave
contribution for the tip. In the MO region of high electron density, the tunneling current
is high (left panel), while the tunneling current is low above the MO nodal
region (right panel). \textbf{(b)} Illustration of a \textit{p}-wave
contribution for the tip. In the MO region of high electron density, the tunneling current
is low (left panel), while the tunneling current is high above the MO nodal
region (right panel).}
\label{fig1}
\end{figure}

We start by comparing the characteristics of \textit{s}- and 
\textit{p}-wave contributions to select an efficient algorithm
(Fig.~\ref{fig1}). For the fully symmetric \textit{s}-wave, significant
tunneling occurs when the tip is placed over regions of high electron
density, while at the MO nodes, the current is substantially reduced due
to quantum destructive interference (Fig.~\ref{fig1}a). In contrast,
\textit{p}-wave tips produce low current in MO regions of high electron
density but high current at the nodes owing to the same mechanism of
quantum interference (Fig.~\ref{fig1}b). It is important to note that
images from a sole \textit{s}-wave tip have generally low image
resolution\cite{gross2011high}, while the commonly used CO-decorated
tips often exhibit comparable \textit{s}- and \textit{p}-wave
contributions\cite{gross2011high,duan2022general,duan2023accurate}.
These observations pose challenge for obtaining images of pristine MOs with
high-resolution. However, the spatially separable nature of \textit{s}-
and \textit{p}-wave signals offers an opportunity to differentiate
themselves using AI algorithms.

\begin{figure}
\centering
\includegraphics[scale=0.75]{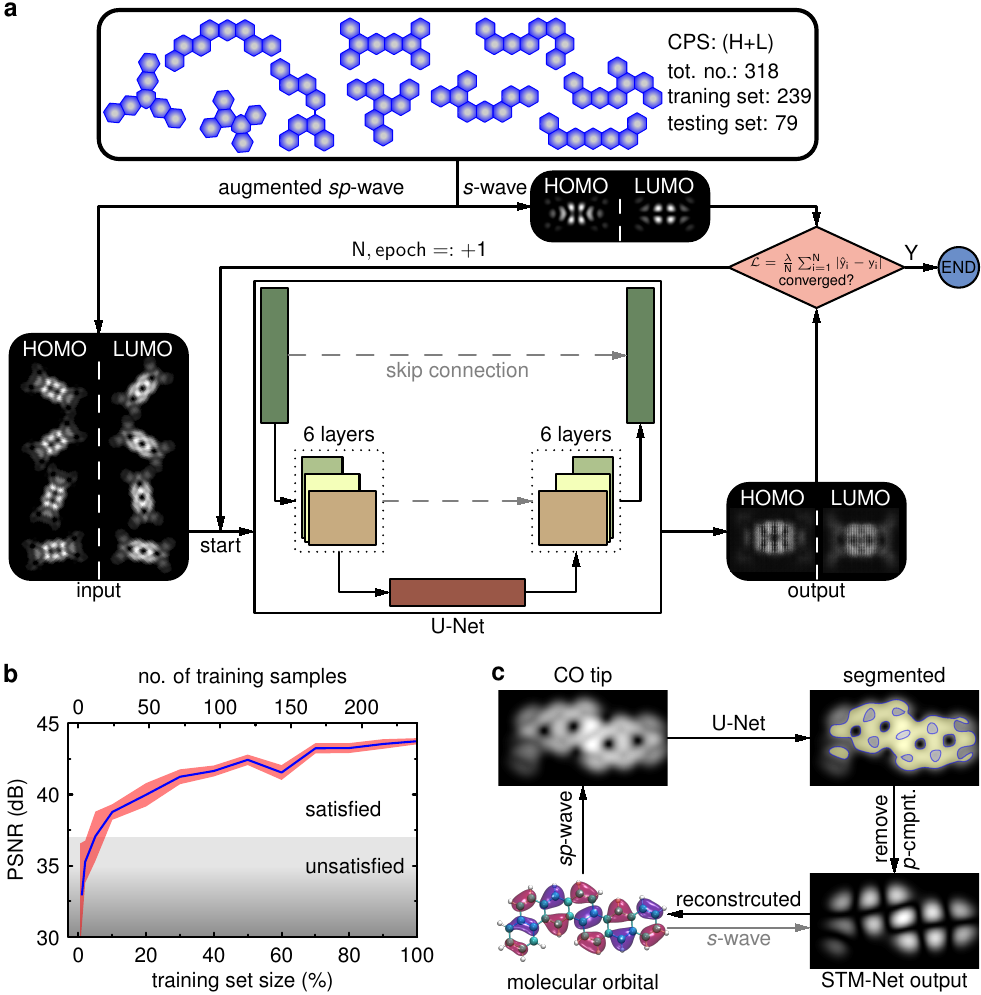}
\caption{\textbf{STM-Net architecture and results}. \textbf{(a)}
Schematic of the training procedure for STM-Net (see details in Methods).
\textbf{(b)} The learning curve for PSNR of STM-Net. The pink-shaded
region represents the 95\% confidence interval obtained from 10 separate
training runs. The blue line shows the mean PSNR values averaged across
these 10 training runs. The gray-shaded region indicates the
performance region of unsatisfied models. \textbf{(c)} Example of MO
reconstruction by STM-Net. Bottom-left: Calculated HOMO of BNT at the DFT
level. Top-left: STM image obtained with the CO-functionalized tip,
which emulates an \textit{sp}-mix wave. Top-right: \textit{p}-wave
contributions identified by STM-Net (highlighted in yellow).
Bottom-right: The output image generated by STM-Net.}
\label{fig2}
\end{figure}

To this end, we constructed a dataset of STM images for the HOMO and
LUMO of 159 representative phenyl-based polycyclic aromatic hydrocarbons
(PAHs) from the COMPAS project\cite{wahab2022compas}, using both pure
metal and CO-functionalized tips (see Methods for details). These
molecules were deliberately selected based on their suitability for
physisorption and the feasibility in experiments via on-surface
synthesis. To avoid the lengthy process of experimental measurements, we
utilized a highly-accurate STM simulation method, which was recently
developed by us using the Bardeen's approximation\cite{bardeen1961prl,duan2022general},
to build the dataset. For our approach, we opted for the U-Net
model\cite{ronneberger2015u}, which is well-suited for the segmentation
task required to distinguish the spatially separable \textit{s}- and
\textit{p}-wave signals. Previous investigations have shown that U-Net
performs well on small datasets\cite{liu2022unet,prasad2021modifying},
making it an ideal choice for our needs.

The architecture of STM-Net is illustrated in Fig.~\ref{fig2}a (see
Methods for more details). To demonstrate the advantage of U-Net on
small datasets, we analyzed the learning curve of STM-Net
(Fig.~\ref{fig2}b). Strikingly, with a small training set, i.e.,
starting from just 10\% of the dataset (corresponding to only 24 images),
STM-Net was able to achieve satisfactory prediction results, with peak
signal-to-noise ratio (PSNR) values above the threshold of 37~dB (see
Methods for PSNR calculation details, as well as Supplementary
Section~1 and Supplementary Fig.~1 for its relationship to the quality
of the output images). With further increases in training set size, the
PSNR of STM-Net eventually exceeded 43~dB (Fig.~\ref{fig2}b), signifying
that the outputs were highly consistent with the ground truth 
\textit{s}-wave images. Similar convergence trends were observed for the
other evaluation metrics (Extended Data Fig.~\ref{fig_ex1}). 

We chose benzo[m]naphtho[2,3-c]tetraphene (BNT) as a representative
molecule to illustrate the working procedure of STM-Net
(Fig.~\ref{fig2}c). The HOMO of the BNT molecule predominantly resides
on two parallel triphenyl units, with a moderate distribution over the
phenyl ring at the lower left corner. When imaging the HOMO of BNT using
a CO-functionalized tip, two band-like patterns with four dark holes
were observed (top-left panel in Fig.~\ref{fig2}c), which were quite
different from the calculated MO as anticipated. Using STM-Net, the
\textit{p}-wave contributions were accurately isolated (yellow shadows
of the top-right panel in Fig.~\ref{fig2}c). By removing the unwanted
\textit{p}-wave signals, the output image from STM-Net (bottom-left
panel in Fig.~\ref{fig2}c) faithfully reproduces all the details of the
corresponding MO. It should be stressed that acquiring a high-resolution
\textit{s}-wave image is inherently challenging in experiments as
previously discussed. Thus, STM-Net introduces the first general machine
learning framework capable of reconstructing pristine MOs from high-resolution
STM images that contain undesired \textit{p}-wave contributions. By leveraging the intrinsic connection between a task in
chemistry and an algorithm in machine learning, STM-Net achieves
remarkable success that is unattainable for traditional image
recognition or transformation algorithms (Supplementary Sections~2-3 and
Supplementary Figs.~2-4).

\subsection*{Application of STM-Net to Experimental Images}
\label{subsec_app}

\begin{figure}
\centering
\includegraphics[scale=0.75]{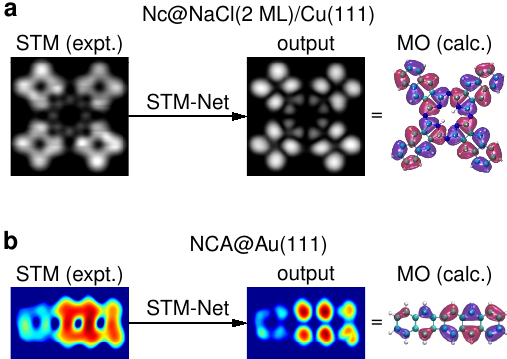}
\caption{\textbf{Application of STM-Net to experimental images}.
\textbf{(a)} Experimentally measured STM image for the HOMO of
naphthalocyanine (Nc) adsorbed on a NaCl(2 ML)/Cu(111) substrate under a
CO-functionalized tip (left, reprinted with permission from
ref.~\citenum{gross2011high}. Copyright 2011 by the American
Physical Society). The output of the STM image after processing by
STM-Net (middle) and the calculated HOMO at the DFT level. \textbf{(b)}
Experimentally measured STM image for the HOMO of an NCA molecule
adsorbed on the Au(111) surface with a CO-functionalized tip (left). The
output of the STM image after processing by STM-Net (middle) and the
calculated HOMO at the DFT level (right).}
\label{fig3}
\end{figure}

The physics-driven nature of STM-Net ensures its strong generalization
ability. To demonstrate this capability, we first applied STM-Net
directly to the experimentally measured STM images for the HOMO of a
naphthalocyanine (Nc) adsorbed on an insulating NaCl thin film covered
Cu(111) surface using a CO-decorated tip\cite{gross2011high}. The NaCl
film acted as a spacer, effectively decoupling the molecule from the
Cu(111) substrate, which created a condition that closely resembled that
in STM-Net. It is noteworthy that Nc contains nitrogen atoms and
heterogeneous pentagonal rings, which significantly differ from the
phenyl-based PAHs in the STM-Net dataset. Despite these differences, the
STM-Net output based on the experimentally measured STM image accurately
reproduces the MO characteristics of Nc (Fig.~\ref{fig3}a).
Specifically, STM-Net not only successfully removed the \textit{p}-wave
contributions from the bright lobe patterns but also effectively
corrected the symmetry of the eight interior patterns (Extended Data
Fig.~\ref{fig_ex2}). This success can be largely attributed to the high
resolution of the original image, facilitated by the decoupling effect
provided by the NaCl film.

Although decoupling the substrate's perturbation on molecular electronic
states can enhance the resolution, introducing a spacer requires extra
experimental effort. In most STM experiments, the standard setup
involves only inert metallic substrates. In this context, metallic
substrates not only serve as supports for molecules but also provide
adatoms to facilitate on-surface synthesis\cite{li2020acsnano}. Thus,
reconstructing pristine MOs from STM images taken on metallic surfaces poses a
more significant challenge for STM-Net. To address this issue, we
examined the product of the debromination cyclization and the subsequent
dimerization reaction of 2,3-bis(dibromomethyl) naphthalene on an Au(111)
substrate, yielding naphtho[2',3':3,4]cyclobuta[1,2-b]anthracene (NCA,
Supplementary Section~4 and Supplementary Fig.~5)\cite{tang2022surface}.
NCA features a four-membered ring, a structure not previously
encountered in the STM-Net dataset. The STM image for the HOMO of NCA
adsorbed on the Au(111) surface is inevitably affected by the metallic
substrate, resulting in considerable distortions (left panel in
Fig.~\ref{fig3}b). Despite these challenges, STM-Net successfully
mitigates much of the noise signals as well as the \textit{p}-wave
contributions (middle panel in Fig.~\ref{fig3}b). As a result, the
STM-Net output not only clearly delineates the critical node information
but also accurately reflects the fact that the MO is concentrated on the
anthracene moiety to the right of the four-membered ring (right panel in
Fig.~\ref{fig3}b).

The above cases not only validate the generalization capability of
STM-Net but also highlight its ability to analyze molecular structures,
offering a new means for identifying products in on-surface synthesis.
Particularly, when nc-AFM fails to deliver conclusive structural
information, high-resolution STM images obtained with functionalized
tips combined with the analysis capabilities of STM-Net present a
powerful alternative for structural analysis.

\subsection*{Adaptation of STM-Net to different tip states}

\begin{figure}
\centering
\includegraphics[scale=0.75]{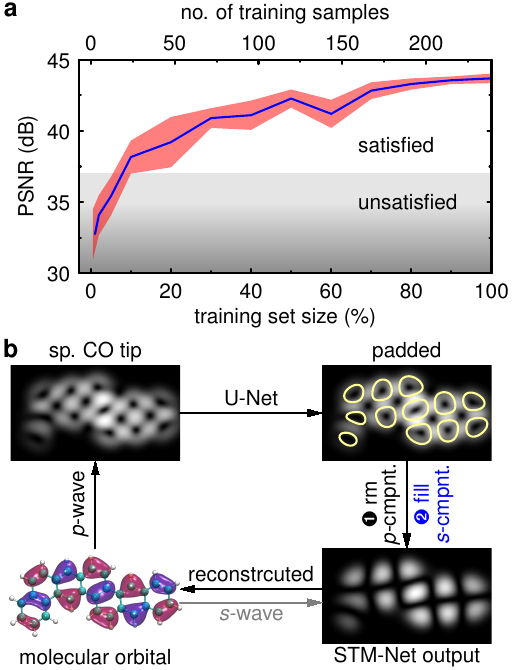}
\caption{\textbf{Adaptation of STM-Net to \textit{p}-to-\textit{s}}.
\textbf{(a)} The learning curve for PSNR of the \textit{p}-to-\textit{s}
STM-Net. The pink-shaded region represents the 95\% confidence interval
obtained from 10 separate training runs. The blue line shows the mean
PSNR values averaged across these 10 training runs. The gray-shaded
region indicates the performance region of unsatisfied models.
\textbf{(b)} Example of MO reconstruction by the \textit{p}-to-\textit{s}
STM-Net. Bottom-left: Calculated HOMO of BNT at the DFT level. Top-left:
STM image obtained with a specially CO-functionalized tip in a pure
\textit{p}-state. Top-right: \textit{s}-wave contributions identified
by the \textit{p}-to-\textit{s} STM-Net (yellow circles). Bottom-right:
The output image generated by the \textit{p}-to-\textit{s} STM-Net.}
\label{fig4}
\end{figure}

The relative ratios of \textit{s}- and \textit{p}-wave contributions for
a CO-decorated tips depend on the interaction between CO and the
metallic tip\cite{gross2011high,mohn2012imaging,duan2023accurate}. The
physics-driven nature of STM-Net enables its adaption to different tip
states. As an extreme condition, we illustrate that the framework can be
adapted for constructing MOs from images acquired using a pure 
\textit{p}-wave tip. The network architecture is kept consistent, with
the only modification being the input used for the training:
specifically replacing \textit{sp}-wave STM images with \textit{p}-wave
STM images (Extended Data Fig.~\ref{fig_ex3}). The PSNR learning curve
exhibits almost identical average behavior compared to the previous
setup, with a converged result exceeding 43~dB again, although the
statistic region of individual learning processes is increased
(Fig.~\ref{fig4}a). This increased variation is also observed in the
other two evaluation metrics, especially when the training size exceeds
30\% (Extended Data Fig.~\ref{fig_ex4}). Such a heightened variation
should be attributed to the extra step required for transforming
\textit{p}-wave characteristics into their \textit{s}-wave equivalents.

To further illustrate the working procedure of STM-Net under these
conditions, we again use the BNT molecule as an example (Fig.~\ref{fig4}b).
The STM image for the HOMO of BNT under a specially CO-decorated tip with pure
\textit{p}-wave state exhibits multiple holes. After training, STM-Net
accurately identifies the regions with these holes that require padding
for reconstructing the \textit{s}-wave image (yellow circles in the
top-right panel of Fig.~\ref{fig4}b). By removing all \textit{p}-wave
signals and subsequently filling in the identified holes, STM-Net
outputs the \textit{s}-wave image (bottom-right panel in Fig.~\ref{fig4}b),
which precisely represents the MO distribution of BNT.

\section*{Conclusion}

MO is a key concept in general chemistry, where directly observing MOs
in real space is a fundamental pursuit of significant scientific
importance. Although considerable progress has been made with the aid of
STM, it remains a substantial challenge, as STM images on the surfaces,
acquired using functional tips, introduce unwanted components other than
the intrinsic MOs of the molecules under study. We addressed this issue
by developing a physics-driven AI network, named STM-Net. This network
effectively eliminates the adverse effects of high-angular-momentum
contributions from functionalized tips by accurately learning to
separate the distinct features originating from the quantum interference,
enabling the accurate reconstruction of frontier MOs. These findings
highlight the principle of using physics-driven AI algorithms to solve
chemical problems, providing a practical route for high-precision MO
imaging. The proposed framework is expected to be extended to diverse
types of decorated tips and more complicated molecular systems,
establishing a research paradigm that closely integrates quantum
chemistry theories with experimental characterization.

\section*{Methods}
\label{sec:method}
\subsection*{Architecture of STM-Net}

The architecture of STM-Net is based on the U-Net
model\cite{ronneberger2015u} with several modifications. For instance,
strided convolution layers have been used in place of pooling layers;
leaky ReLU activation functions were used in the down-sampling layers to
prevent gradient vanishing; and transposed convolutions were employed
instead of upsampling (interpolation) followed by a convolutional layer.
The code of STM-Net has been made publicly available\cite{stm_git}. 

The specific U-Net encoder and decoder in STM-Net were structured as
follows:

-\textbf{U-net encoder}: $\text{C}_{64}$–$\text{C}_{128}$–$\text{C}_{256}$–$\text{C}_{512}$–$\text{C}_{512}$–$\text{C}_{512}$–$\text{C}_{512}$

-\textbf{U-Net Decoder}: $\text{CD}_{512}$–$\text{CD}_{1024}$–$\text{C}_{1024}$–$\text{C}_{1024}$–$\text{C}_{512}$–$\text{C}_{256}$–$\text{C}_{128}$\\
where $\text{C}_i$ represents a convolution-batch normalization-ReLU
layer with $i$ filters and $\text{CD}_j$ indicates a convolution-batch
normalization-Dropout-ReLU layer with a 50\% dropout rate (Supplementary
Section~5 and Supplementary Figs.~6-7). All convolution layers utilized
a $4\times4$ spatial filter with a stride of 2. This configuration
implies that features were down-sampled and up-sampled by a factor of 2
in the encoder and decoder.

A convolution layer was attached to the final decoder layer to map the
output to the required number of channels (1 in the present work),
followed by a Tanh activation function as
\begin{equation}
\text{Tanh}(x) = \frac{e^x-e^{-x}}{e^x + e^{-x}}.
\label{eq_ftanh}
\end{equation}
All ReLU activation functions in the encoder were replaced by leaky
ReLU, defined as
\begin{equation}
\text{LeakyReLU}(x) = \max(0, x) + 0.2 \min(0, x).
\label{eq_flklu}
\end{equation}
In the decoder, the ReLU activation function
\begin{equation}
\text{ReLU}(x) = \max(0, x)
\label{eq_flu}
\end{equation}
was used. The U-Net architecture includes skip connections that link the
activation output of each encoder layer $i$ to the corresponding decoder
layer $n-i$, where $n$ is the total number of layers. These skip
connections double the number of channels in the decoder compared to the
encoder. Batch normalization is not applied to the first $\text{C}_{64}$
layer in the encoder.

\subsection*{Training parameters of STM-Net}

We first translated each STM image to focus on the central region and
then resized the image to $286\times286$ pixels. Subsequently, random
cropping was applied to further adjust the images to a final size of
$256\times256$ pixels. This approach prevents the model from
over-relying on the fixed positions of objects in the images. The
process of jittering with some further random mirroring also enhances
the model's robustness against minor variations, improving its
generalization capability. All networks were trained from scratch, with
weights initialized from a Gaussian distribution with a mean of 0 and a
standard deviation of 0.02.

The dataset  consisted of 318 theoretically simulated STM images
generated at the density functional theory (DFT) level (section
``Dataset in STM-Net'' in Methods) and was trained over 1000 epochs with
a batch size of 16. The dataset was randomly divided into training and
testing sets in a 75:25 ratio (Supplementary Figs.~8-15). During the
first 500 epochs, the learning rate was set to 0.0002, and for the
remaining 500 epochs, it linearly decayed to 0. The training process
utilized the loss function as:
\begin{equation}
\mathcal{L}_1=\frac{1}{N\cdot{C}\cdot{H}\cdot{W}}
\sum_{n=1}^N\sum_{c=1}^C\sum_{h=1}^H\sum_{w=1}^W
\left|I_{n,c}^{\text{out}}(h,w)-I_{n,c}^{\text{gt}}(h,w)\right|,
\label{eq_l1}
\end{equation}
where $N$ is the batch size, $C$ is the number of image channels, $H$
($W$) is the height (width) of the images, $I^{\text{out}}$ is the
output image of STM-Net, and $I^{\text{gt}}$ is the ground truth of the
\textit{s}-wave STM image. In this work, we set $N=16$, $C=1$, $H=256$,
and $W=256$. The Adam optimizer with $\beta_1=0.5$ and $\beta_2=0.999$
was used for back-prorogation, To process the real experimental
measurements using STM-Net, the dataset was augmented by rotating each
image in every 5$^\circ$ increment as illustrated in Fig.~\ref{fig2}a
and Extended Data Fig.~\ref{fig_ex3}. This augmentation expands the
dataset by a factor of 72.

\subsection*{Dataset in STM-Net}

We selected 159 representative hexagonal PAHs from the COMPAS
project\cite{wahab2022compas} database (Supplementary Section~6). The
molecular structures were first optimized in the gas phase using DFT at
the X3LYP\cite{xu2004x3lyp} level with the cc-pVTZ basis
set\cite{dunning1989gaussian,peterson1994benchmark,woon1993gaussian,kendall1992electron},
as implemented in the Gaussian 16 suite of programs\cite{g16}. This
setting corresponds to isolated molecular systems on insulating
substrates in STM measurements. After obtaining the wavefunctions of the
isolated molecules, the HOMO and LUMO STM images were simulated using
Bardeen’s approximation\cite{bardeen1961prl,duan2022general}.
Specifically, the tunneling current in atomic units is given by
\begin{equation}
I=2\pi\int_{-\infty}^\infty[f_t(E-eV)-f_s(E)]\rho_t(E-eV)\rho_s(E)
|M_{st}|^2dE,
\label{eq_bardeen}
\end{equation}
where subscripts ``$t$'' and ``$s$'' refer to functions associated with
the tip and substrate, $f$ stands for the Fermi-Dirac distribution, $e$
is the elementary charge, $V$ is the applied bias voltage, $\rho$ is the
density of states, and $M_{st}$ is the transfer Hamiltonian matrix
element. The energy reference levels for the tip and substrate are set
to the corresponding Fermi levels ($E_F$) under the applied bias.
High-resolution STM measurements were typically performed at low
temperatures. For isolated molecules, $\rho_s$ becomes a Dirac
$\delta$-function, representing discrete states. In this scenario, the
tunneling current can be simplified as
\begin{equation}
I=2\pi\sum_{E_s=E_F}^{E_F+eV}\rho_t(E_s-eV)|M_{st}|^2
\label{eq_bardeen-s}
\end{equation}
where $E_s$ denotes the molecular energy levels. In a practical STM
image simulation, $\rho_t$ was assumed to be a constant.

The transfer Hamiltonian matrix can be calculated
as\cite{bardeen1961prl,duan2022general}
\begin{equation}
M_{st}=-\frac{1}{2}\int_{\Sigma}
\left(\Psi_t^\ast\nabla\Psi_s-\Psi_s^\ast\nabla\Psi_t\right)\cdot\text{d}\mathbf{s}
\label{eq_mst}
\end{equation}
Here, $\Psi_t$ and $\Psi_s$ denote the quasi-particle wavefunctions of
the tip and sample, respectively, $\Sigma$ represents the separation
surface between the tip and sample regions, and $d\mathbf{s}$ is the
surface element vector on $\Sigma$. In our calculations, \textit{s}-
and \textit{p}-type Gaussian functions with an exponent of
$\alpha=0.25~\text{Bohr}^{-2}$ were used to mimic the \textit{s}- and
\textit{p}-wave tips, respectively. For the \textit{sp}-mixed tip, the
contributions of \textit{s}- and \textit{p}-wave were set to 1:1 and
$z_0$ was set to 0.1~\AA{} above the molecules to mimic the behavior of
a typical CO-functionalized STM tip\cite{gross2011high,duan2022general,duan2023accurate}.

\subsection*{Evaluation metrics for STM-Net}

We evaluated the outputs of STM-Net using widely-recognized metrics,
namely PSNR, structural similarity index measure (SSIM), and learned
perceptual image patch similarity (LPIPS). Given that these metrics
exhibited generally consistent trends, PSNR, which is straightforward
and widely applicable, was chosen in the main text. 

PSNR\cite{korhonen2012peak} measures the ratio between the maximum
possible value of a signal and the degradation caused by a noise,
providing an objective quantification of image restoration degree by
assessing the mean squared error (MSE). The definition of PSNR is
\begin{equation}
\text{PSNR}=20\log_{10}
\left(\frac{\text{MAX}_f}{\sqrt{\text{MSE}}}\right),
\label{eq_psnr}
\end{equation}
where $\text{MAX}_f$ represents the maximum possible pixel value of the
image, and MSE can be calculated as
\begin{equation}
\text{MSE}=\frac{1}{H\cdot{W}}\sum_{h=1}^{H}\sum_{w=1}^{W}
\|I^{\text{out}}(h,w)-I^{\text{gt}}(h,w)\|^2.
\label{eq_mse}
\end{equation}
The logarithmic decibel scale used in PSNR calculation is particularly
suitable for evaluating signals that have a broad dynamic range. 

SSIM\cite{wang2004image} evaluates image quality from a perceptual
perspective, considering degradation in structural information and
incorporating the effects of luminance and contrast masking. Unlike PSNR,
which focuses on pixel-wise differences, SSIM emphasizes changes in
structural information and calculates a composite similarity score that
reflects these aspects along with luminance and contrast (see details in
Supplementary Section~7.1).

LPIPS\cite{zhang2018unreasonable} uses deep learning techniques to
evaluate visual perceptual similarity. Through pre-trained convolutional
neural networks, it extracts high-level features from images to
calculate perceptual similarity score. This metrics especially suitable
for tasks like generative model evaluation or those requiring high
perceptual accuracy (see details in Supplementary Section~7.2).

\subsection*{Experimental STM measurements}

The STM images of NCA were obtained at 4.6~K with a commercial LT-STM
(Scienta Omicron). The single crystal Au(111) was supplied by MaTeck
GmbH and was cleaned through cyclic Ar-ion sputtering and annealing. The
precursors (2,3-bis(dibromomethyl)naphthalene)\cite{tang2022surface}
were deposited onto the Au(111) surfaces held at room-temperature by
organic molecular beam deposition technique from a commercial evaporator
(Kentax GmbH). The sample of NCA was prepared by on-surface reactions of
the precursors (thermal annealing at 400~K for 30~min). To minimize the
interference of the Au(111) substrate, the $dI/dV$ mapping of NCA was
conducted in the constant-height mode by using a CO-functionalized tip
with a resonance frequency of $f_0\approx27.5~\text{kHz}$ and a quality
factor $Q>10^4$. All STM images were observed under ultra-high vacuum
conditions with a base pressure better than $1.0\times10^{-10}$~mbar.


\bibliography{STMNet}% common bib file

\begin{thebibliography}{10}
\expandafter\ifx\csname url\endcsname\relax
  \def\url#1{\burl{#1}}\fi
\expandafter\ifx\csname urlprefix\endcsname\relax\def\urlprefix{URL }\fi
\providecommand{\bibinfo}[2]{#2}
\providecommand{\eprint}[2][]{\url{#2}}
\providecommand{\doi}[1]{\url{https://doi.org/#1}}
\bibcommenthead

\bibitem{mulliken1932electronic}
\bibinfo{author}{Mulliken, R.~S.}
\newblock \bibinfo{title}{Electronic structures of polyatomic molecules and
  valence. {II}. {General} considerations}.
\newblock \emph{\bibinfo{journal}{Phys. Rev.}} \textbf{\bibinfo{volume}{41}},
  \bibinfo{pages}{49} (\bibinfo{year}{1932}).

\bibitem{fukui1952molecular}
\bibinfo{author}{Fukui, K.}, \bibinfo{author}{Yonezawa, T.} \&
  \bibinfo{author}{Shingu, H.}
\newblock \bibinfo{title}{A molecular orbital theory of reactivity in aromatic
  hydrocarbons}.
\newblock \emph{\bibinfo{journal}{J. Chem. Phys.}}
  \textbf{\bibinfo{volume}{20}}, \bibinfo{pages}{722--725}
  (\bibinfo{year}{1952}).

\bibitem{woodward1965jacs}
\bibinfo{author}{Woodward, R.~B.} \& \bibinfo{author}{Roald, H.}
\newblock \bibinfo{title}{Stereochemistry of electrocyclic reactions}.
\newblock \emph{\bibinfo{journal}{J. Am. Chem. Soc.}}
  \textbf{\bibinfo{volume}{87}}, \bibinfo{pages}{395--397}
  (\bibinfo{year}{1965}).

\bibitem{mulder2011orbitals}
\bibinfo{author}{Mulder, P.}
\newblock \bibinfo{title}{Are orbitals observable?}
\newblock \emph{\bibinfo{journal}{Int. J. Philos. Chem.}}
  \textbf{\bibinfo{volume}{17}}, \bibinfo{pages}{24--35}
  (\bibinfo{year}{2011}).

\bibitem{pham2017can}
\bibinfo{author}{Pham, B.~Q.} \& \bibinfo{author}{Gordon, M.~S.}
\newblock \bibinfo{title}{Can orbitals really be observed in scanning tunneling
  microscopy experiments?}
\newblock \emph{\bibinfo{journal}{J. Phys. Chem. A}}
  \textbf{\bibinfo{volume}{121}}, \bibinfo{pages}{4851--4852}
  (\bibinfo{year}{2017}).

\bibitem{zuo1999direct}
\bibinfo{author}{Zuo, J.-M.}, \bibinfo{author}{Kim, M.},
  \bibinfo{author}{O'keeffe, M.} \& \bibinfo{author}{Spence, J.}
\newblock \bibinfo{title}{Direct observation of \textit{d}-orbital holes and
  {Cu--Cu} bonding in {Cu$_2$O}}.
\newblock \emph{\bibinfo{journal}{Nature}} \textbf{\bibinfo{volume}{401}},
  \bibinfo{pages}{49--52} (\bibinfo{year}{1999}).

\bibitem{repp2005molecules}
\bibinfo{author}{Repp, J.}, \bibinfo{author}{Meyer, G.},
  \bibinfo{author}{Stojkovi{\'c}, S.~M.}, \bibinfo{author}{Gourdon, A.} \&
  \bibinfo{author}{Joachim, C.}
\newblock \bibinfo{title}{Molecules on insulating films: Scanning-tunneling
  microscopy imaging of individual molecular orbitals}.
\newblock \emph{\bibinfo{journal}{Phys. Rev. Lett.}}
  \textbf{\bibinfo{volume}{94}}, \bibinfo{pages}{026803}
  (\bibinfo{year}{2005}).

\bibitem{gross2011recent}
\bibinfo{author}{Gross, L.}
\newblock \bibinfo{title}{Recent advances in submolecular resolution with
  scanning probe microscopy}.
\newblock \emph{\bibinfo{journal}{Nat. Chem.}} \textbf{\bibinfo{volume}{3}},
  \bibinfo{pages}{273--278} (\bibinfo{year}{2011}).

\bibitem{kitou2017successive}
\bibinfo{author}{Kitou, S.} \emph{et~al.}
\newblock \bibinfo{title}{Successive dimensional transition in
  {(TMTTF)$_2$PF$_6$} revealed by synchrotron {X}-ray diffraction}.
\newblock \emph{\bibinfo{journal}{Phy. Rev. Lett.}}
  \textbf{\bibinfo{volume}{119}}, \bibinfo{pages}{065701}
  (\bibinfo{year}{2017}).

\bibitem{tersoff1985theory}
\bibinfo{author}{Tersoff, J.} \& \bibinfo{author}{Hamann, D.~R.}
\newblock \bibinfo{title}{Theory of the scanning tunneling microscope}.
\newblock \emph{\bibinfo{journal}{Phy. Rev. B}} \textbf{\bibinfo{volume}{31}},
  \bibinfo{pages}{805} (\bibinfo{year}{1985}).

\bibitem{chen2021introduction}
\bibinfo{author}{Chen, C.~J.}
\newblock \emph{\bibinfo{title}{Introduction to Scanning Tunneling Microscopy}}
  \bibinfo{edition}{3} edn (\bibinfo{publisher}{Oxford University Press, USA},
  \bibinfo{year}{2021}).

\bibitem{duan2015jacs}
\bibinfo{author}{Duan, S.} \emph{et~al.}
\newblock \bibinfo{title}{Theoretical modeling of plasmon-enhanced raman images
  of a single molecule with subnanometer resolution}.
\newblock \emph{\bibinfo{journal}{J. Am. Chem. Soc.}}
  \textbf{\bibinfo{volume}{137}}, \bibinfo{pages}{9515--9518}
  (\bibinfo{year}{2015}).

\bibitem{duan2020jacs}
\bibinfo{author}{Duan, S.}, \bibinfo{author}{Zhang, I.~Y.},
  \bibinfo{author}{Xie, Z.} \& \bibinfo{author}{Xu, X.}
\newblock \bibinfo{title}{Identification of water hexamer on cu(111) surfaces}.
\newblock \emph{\bibinfo{journal}{J. Am. Chem. Soc.}}
  \textbf{\bibinfo{volume}{142}}, \bibinfo{pages}{6902--6906}
  (\bibinfo{year}{2020}).

\bibitem{gross2011high}
\bibinfo{author}{Gross, L.} \emph{et~al.}
\newblock \bibinfo{title}{High-resolution molecular orbital imaging using a
  \textit{p}-wave {STM} tip}.
\newblock \emph{\bibinfo{journal}{Phys. Rev. Lett.}}
  \textbf{\bibinfo{volume}{107}}, \bibinfo{pages}{086101}
  (\bibinfo{year}{2011}).

\bibitem{chen1990tunneling}
\bibinfo{author}{Chen, C.~J.}
\newblock \bibinfo{title}{Tunneling matrix elements in three-dimensional space:
  The derivative rule and the sum rule}.
\newblock \emph{\bibinfo{journal}{Phys. Rev. B}} \textbf{\bibinfo{volume}{42}},
  \bibinfo{pages}{8841} (\bibinfo{year}{1990}).

\bibitem{duan2022general}
\bibinfo{author}{Duan, S.}, \bibinfo{author}{Tian, G.} \& \bibinfo{author}{Xu,
  X.}
\newblock \bibinfo{title}{A general framework of scanning tunneling microscopy
  based on {B}ardeen's approximation for isolated molecules}.
\newblock \emph{\bibinfo{journal}{JACS Au}} \textbf{\bibinfo{volume}{3}},
  \bibinfo{pages}{86--92} (\bibinfo{year}{2022}).

\bibitem{duan2023accurate}
\bibinfo{author}{Duan, S.} \& \bibinfo{author}{Xu, X.}
\newblock \bibinfo{title}{Accurate simulations of scanning tunneling
  microscope: Both tip and substrate states matter}.
\newblock \emph{\bibinfo{journal}{J. Phys. Chem. Lett.}}
  \textbf{\bibinfo{volume}{14}}, \bibinfo{pages}{6726--6735}
  (\bibinfo{year}{2023}).

\bibitem{beyan2023review}
\bibinfo{author}{Beyan, E. V.~P.}, \bibinfo{author}{Rossy, A. G.~C.}
  \emph{et~al.}
\newblock \bibinfo{title}{A review of {AI} image generator: influences,
  challenges, and future prospects for architectural field}.
\newblock \emph{\bibinfo{journal}{J. Artif. Intell. Archt.}}
  \textbf{\bibinfo{volume}{2}}, \bibinfo{pages}{53--65} (\bibinfo{year}{2023}).

\bibitem{castiglioni2021ai}
\bibinfo{author}{Castiglioni, I.} \emph{et~al.}
\newblock \bibinfo{title}{{AI} applications to medical images: From machine
  learning to deep learning}.
\newblock \emph{\bibinfo{journal}{Phys. Medica}} \textbf{\bibinfo{volume}{83}},
  \bibinfo{pages}{9--24} (\bibinfo{year}{2021}).

\bibitem{krizhevsky2012nips}
\bibinfo{author}{Krizhevsky, A.}, \bibinfo{author}{Sutskever, I.} \&
  \bibinfo{author}{Hinton, G.~E.}
  \emph{\bibinfo{title}{{ImageNet} classification with deep convolutional
  neural networks}}.
\newblock (eds \bibinfo{editor}{Pereira, F.}, \bibinfo{editor}{Burges, C.~J.},
  \bibinfo{editor}{Bottou, L.} \& \bibinfo{editor}{Weinberger, K.~Q.})
  \emph{\bibinfo{booktitle}{Advances in Neural Information Processing
  Systems}}, Vol.~\bibinfo{volume}{25} (\bibinfo{publisher}{Curran Associates,
  Inc.}, \bibinfo{year}{2012}).

\bibitem{yang2024mattersim}
\bibinfo{author}{Yang, H.} \emph{et~al.}
\newblock \bibinfo{title}{{MatterSim}: A deep learning atomistic model across
  elements, temperatures and pressures}.
\newblock \emph{\bibinfo{journal}{arXiv preprint arXiv:2405.04967}}
  (\bibinfo{year}{2024}).

\bibitem{alberts2023large}
\bibinfo{author}{Alberts, I.~L.} \emph{et~al.}
\newblock \bibinfo{title}{Large language models ({LLM}) and {ChatGPT}: What
  will the impact on nuclear medicine be?}
\newblock \emph{\bibinfo{journal}{Eur. J. Nucl. Med. Mol. Imaging}}
  \textbf{\bibinfo{volume}{50}}, \bibinfo{pages}{1549--1552}
  (\bibinfo{year}{2023}).

\bibitem{ronneberger2015u}
\bibinfo{author}{Ronneberger, O.}, \bibinfo{author}{Fischer, P.} \&
  \bibinfo{author}{Brox, T.}
  \emph{\bibinfo{title}{U-{N}et: Convolutional networks for biomedical image
  segmentation}}.
\newblock (eds \bibinfo{editor}{Navab, N.}, \bibinfo{editor}{Hornegger, J.},
  \bibinfo{editor}{Wells, W.~M.} \& \bibinfo{editor}{Frangi, A.~F.})
  \emph{\bibinfo{booktitle}{Medical Image Computing and Computer-Assisted
  Intervention -- MICCAI}}, \bibinfo{pages}{234--241}
  (\bibinfo{publisher}{Springer International Publishing},
  \bibinfo{address}{Cham}, \bibinfo{year}{2015}).

\bibitem{goodfellow2020generative}
\bibinfo{author}{Goodfellow, I.} \emph{et~al.}
\newblock \bibinfo{title}{Generative adversarial networks}.
\newblock \emph{\bibinfo{journal}{Commun. ACM}} \textbf{\bibinfo{volume}{63}},
  \bibinfo{pages}{139--144} (\bibinfo{year}{2020}).

\bibitem{ho2020denoising}
\bibinfo{author}{Ho, J.}, \bibinfo{author}{Jain, A.} \&
  \bibinfo{author}{Abbeel, P.}
\newblock \bibinfo{title}{Denoising diffusion probabilistic models}.
\newblock \emph{\bibinfo{journal}{Adv. Neural Inf. Process. Syst.}}
  \textbf{\bibinfo{volume}{33}}, \bibinfo{pages}{6840--6851}
  (\bibinfo{year}{2020}).

\bibitem{krull2020artificial}
\bibinfo{author}{Krull, A.}, \bibinfo{author}{Hirsch, P.},
  \bibinfo{author}{Rother, C.}, \bibinfo{author}{Schiffrin, A.} \&
  \bibinfo{author}{Krull, C.}
\newblock \bibinfo{title}{Artificial-intelligence-driven scanning probe
  microscopy}.
\newblock \emph{\bibinfo{journal}{Commun. Phys.}} \textbf{\bibinfo{volume}{3}},
  \bibinfo{pages}{54} (\bibinfo{year}{2020}).

\bibitem{su2024natsyn}
\bibinfo{author}{Su, J.} \emph{et~al.}
\newblock \bibinfo{title}{Intelligent synthesis of magnetic nanographenes via
  chemist-intuited atomic robotic probe}.
\newblock \emph{\bibinfo{journal}{Nat. Synth.}} \textbf{\bibinfo{volume}{3}},
  \bibinfo{pages}{466--476} (\bibinfo{year}{2024}).

\bibitem{zhu2024jacs}
\bibinfo{author}{Zhu, Z.} \emph{et~al.}
\newblock \bibinfo{title}{Autonomous scanning tunneling microscopy imaging via
  deep learning}.
\newblock \emph{\bibinfo{journal}{J. Am. Chem. Soc.}}
  \textbf{\bibinfo{volume}{146}}, \bibinfo{pages}{29199--29206}
  (\bibinfo{year}{2024}).

\bibitem{wahab2022compas}
\bibinfo{author}{Wahab, A.}, \bibinfo{author}{Pfuderer, L.},
  \bibinfo{author}{Paenurk, E.} \& \bibinfo{author}{Gershoni-Poranne, R.}
\newblock \bibinfo{title}{The {COMPAS} project: A computational database of
  polycyclic aromatic systems. {P}hase 1: \textit{cata}-{C}ondensed
  polybenzenoid hydrocarbons}.
\newblock \emph{\bibinfo{journal}{J. Chem. Inf. Model.}}
  \textbf{\bibinfo{volume}{62}}, \bibinfo{pages}{3704--3713}
  (\bibinfo{year}{2022}).

\bibitem{bardeen1961prl}
\bibinfo{author}{Bardeen, J.}
\newblock \bibinfo{title}{Tunnelling from a many-particle point of view}.
\newblock \emph{\bibinfo{journal}{Phys. Rev. Lett.}}
  \textbf{\bibinfo{volume}{6}}, \bibinfo{pages}{57--59} (\bibinfo{year}{1961}).

\bibitem{liu2022unet}
\bibinfo{author}{Liu, F.} \& \bibinfo{author}{Wang, L.}
\newblock \bibinfo{title}{{UNet}-based model for crack detection integrating
  visual explanations}.
\newblock \emph{\bibinfo{journal}{Constr. Build. Mater.}}
  \textbf{\bibinfo{volume}{322}}, \bibinfo{pages}{126265}
  (\bibinfo{year}{2022}).

\bibitem{prasad2021modifying}
\bibinfo{author}{Prasad, P. J.~R.}, \bibinfo{author}{Elle, O.~J.},
  \bibinfo{author}{Lindseth, F.}, \bibinfo{author}{Albregtsen, F.} \&
  \bibinfo{author}{Kumar, R.~P.}
  \emph{\bibinfo{title}{{Modifying {U-Net} for small dataset: A
  simplified {U-Net} version for liver parenchyma segmentation}}}.
\newblock (eds \bibinfo{editor}{Mazurowski, M.~A.} \& \bibinfo{editor}{Drukker,
  K.}) \emph{\bibinfo{booktitle}{Medical Imaging 2021: Computer-Aided
  Diagnosis}}, Vol. \bibinfo{volume}{11597}, \bibinfo{pages}{115971O}.
  \bibinfo{organization}{International Society for Optics and Photonics}
  (\bibinfo{publisher}{SPIE}, \bibinfo{year}{2021}).

\bibitem{li2020acsnano}
\bibinfo{author}{Li, S.} \emph{et~al.}
\newblock \bibinfo{title}{Structural phase transitions of molecular
  self-assembly driven by nonbonded metal adatoms}.
\newblock \emph{\bibinfo{journal}{ACS Nano}} \textbf{\bibinfo{volume}{14}},
  \bibinfo{pages}{6331--6338} (\bibinfo{year}{2020}).

\bibitem{tang2022surface}
\bibinfo{author}{Tang, Y.} \emph{et~al.}
\newblock \bibinfo{title}{On-surface debromination of 2,3-bis(dibromomethyl)-
  and 2,3-bis(bromomethyl) naphthalene: Dimerization or polymerization?}
\newblock \emph{\bibinfo{journal}{Angew. Chem. Int. Ed.}}
  \textbf{\bibinfo{volume}{61}}, \bibinfo{pages}{e202204123}
  (\bibinfo{year}{2022}).

\bibitem{mohn2012imaging}
\bibinfo{author}{Mohn, F.}, \bibinfo{author}{Gross, L.}, \bibinfo{author}{Moll,
  N.} \& \bibinfo{author}{Meyer, G.}
\newblock \bibinfo{title}{Imaging the charge distribution within a single
  molecule}.
\newblock \emph{\bibinfo{journal}{Nat. Nanotechnol.}}
  \textbf{\bibinfo{volume}{7}}, \bibinfo{pages}{227--231}
  (\bibinfo{year}{2012}).

\bibitem{stm_git}
\bibinfo{title}{{STM-Net}: A physics-driven deep learning framework for
  constructing molecular orbitals from {STM} images}.
\newblock \bibinfo{howpublished}{\url{https://github.com/ZeHeru/STM_net}}.

\bibitem{xu2004x3lyp}
\bibinfo{author}{Xu, X.} \& \bibinfo{author}{Goddard~III, W.~A.}
\newblock \bibinfo{title}{The {X3LYP} extended density functional for accurate
  descriptions of nonbond interactions, spin states, and thermochemical
  properties}.
\newblock \emph{\bibinfo{journal}{Proc. Natl. Acad. Sci. USA}}
  \textbf{\bibinfo{volume}{101}}, \bibinfo{pages}{2673--2677}
  (\bibinfo{year}{2004}).

\bibitem{dunning1989gaussian}
\bibinfo{author}{Dunning~Jr, T.~H.}
\newblock \bibinfo{title}{Gaussian basis sets for use in correlated molecular
  calculations. {I}. {The} atoms boron through neon and hydrogen}.
\newblock \emph{\bibinfo{journal}{J. Chem. Phys.}}
  \textbf{\bibinfo{volume}{90}}, \bibinfo{pages}{1007--1023}
  (\bibinfo{year}{1989}).

\bibitem{peterson1994benchmark}
\bibinfo{author}{Peterson, K.~A.}, \bibinfo{author}{Woon, D.~E.} \&
  \bibinfo{author}{Dunning~Jr, T.~H.}
\newblock \bibinfo{title}{Benchmark calculations with correlated molecular wave
  functions. {IV}. {The} classical barrier height of the
  {H}+{H$_2$}$\to${H$_2$}+{H} reaction}.
\newblock \emph{\bibinfo{journal}{J. Chem. Phys.}}
  \textbf{\bibinfo{volume}{100}}, \bibinfo{pages}{7410--7415}
  (\bibinfo{year}{1994}).

\bibitem{woon1993gaussian}
\bibinfo{author}{Woon, D.~E.} \& \bibinfo{author}{Dunning~Jr, T.~H.}
\newblock \bibinfo{title}{Gaussian basis sets for use in correlated molecular
  calculations. {III}. {The} atoms aluminum through argon}.
\newblock \emph{\bibinfo{journal}{J. Chem. Phys.}}
  \textbf{\bibinfo{volume}{98}}, \bibinfo{pages}{1358--1371}
  (\bibinfo{year}{1993}).

\bibitem{kendall1992electron}
\bibinfo{author}{Kendall, R.~A.}, \bibinfo{author}{Dunning~Jr, T.~H.} \&
  \bibinfo{author}{Harrison, R.~J.}
\newblock \bibinfo{title}{Electron affinities of the first-row atoms revisited.
  {S}ystematic basis sets and wave functions}.
\newblock \emph{\bibinfo{journal}{J. Chem. Phys.}}
  \textbf{\bibinfo{volume}{96}}, \bibinfo{pages}{6796--6806}
  (\bibinfo{year}{1992}).

\bibitem{g16}
\bibinfo{author}{Frisch, M.~J.} \emph{et~al.}
\newblock \bibinfo{title}{Gaussian 16 {R}evision {C}.01}
  (\bibinfo{year}{2016}).
\newblock \bibinfo{note}{Gaussian Inc. Wallingford CT}.

\bibitem{korhonen2012peak}
\bibinfo{author}{Korhonen, J.} \& \bibinfo{author}{You, J.}
\newblock \bibinfo{editor}{Burnett, I.} (ed.) \emph{\bibinfo{title}{Peak
  signal-to-noise ratio revisited: Is simple beautiful?}}
\newblock (ed.\bibinfo{editor}{Burnett, I.}) \emph{\bibinfo{booktitle}{Fourth
  International Workshop on Quality of Multimedia Experience}},
  \bibinfo{pages}{37--38} (\bibinfo{year}{2012}).

\bibitem{wang2004image}
\bibinfo{author}{Wang, Z.}, \bibinfo{author}{Bovik, A.~C.},
  \bibinfo{author}{Sheikh, H.~R.} \& \bibinfo{author}{Simoncelli, E.~P.}
\newblock \bibinfo{title}{Image quality assessment: from error visibility to
  structural similarity}.
\newblock \emph{\bibinfo{journal}{IEEE Trans. Image Process.}}
  \textbf{\bibinfo{volume}{13}}, \bibinfo{pages}{600--612}
  (\bibinfo{year}{2004}).

\bibitem{zhang2018unreasonable}
\bibinfo{author}{Zhang, R.}, \bibinfo{author}{Isola, P.},
  \bibinfo{author}{Efros, A.~A.}, \bibinfo{author}{Shechtman, E.} \&
  \bibinfo{author}{Wang, O.}
  \emph{\bibinfo{title}{The unreasonable effectiveness of deep features as a
  perceptual metric}}.
\newblock (eds \bibinfo{editor}{Forsyth, D.}, \bibinfo{editor}{Laptev, I.},
  \bibinfo{editor}{Oliva, A.} \& \bibinfo{editor}{Ramanan, D.})
  \emph{\bibinfo{booktitle}{{IEEE/CVF Conference on Computer Vision and Pattern
  Recognition (CVPR)}}}, \bibinfo{pages}{586--595} (\bibinfo{year}{2018}).

\end{thebibliography}

%%==========================================%%
%% Additional information after references  %%
%%==========================================%%

\section*{Acknowledgements}
This work was supported by the National Key R\&D Program of China
(2024YFA1208104), the National Natural Science Foundation of China (Nos.
22393911, 22473028, U24A20496, and 22472114), and the Innovation Program
for Quantum Science and Technology (Nos. 2021ZD0303301 and 2021ZD0303305).

\section*{Author contributions}
X.X. and S.D. conceived the research. X.X., S.D., H.Z. and L.C. jointly
supervised the work. Y.Z. performed the simulations. R.X. performed the
experiments. X.X., S.D., H.Z., Y.Z. and R.X. interpreted the data. All
authors contributed to the writing of the manuscript.

\section*{Competing interests}
The authors declare no competing interests.

\renewcommand{\theHfigure}{A.Abb.\arabic{figure}}
\setcounter{figure}{0}
\renewcommand{\figurename}{Extended Data Fig.}

\clearpage
\newpage

\begin{figure}
\centering
\includegraphics[scale=0.75]{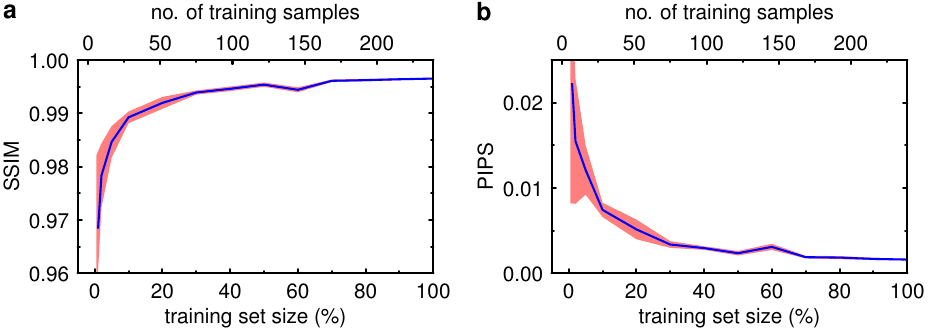}
\caption{\textbf{Other metrics for STM-Net}. The learning curves for
SSIM (\textbf{a}) and LPIPS (\textbf{b}) of STM-Net. The pink-shaded
region represents the 95\% confidence interval obtained from 10 separate
training runs. The blue line shows the mean metrics values averaged
across these 10 training runs.}
\label{fig_ex1}
\end{figure}

\clearpage
\newpage

\begin{figure}
\centering
\includegraphics[scale=0.75]{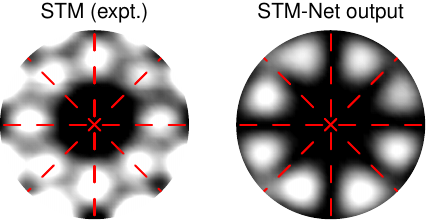}
\caption{\textbf{Zoomed and rescaled Nc images}. Zoomed and rescaled
version of the experimentally measured STM image for the HOMO of
Nc@NaCl(2~ML)@Cu(111) (left) and its corresponding STM-Net output (right).
The red dashed lines indicate the nodal planes owing to the $a_{\text{u}}$
irreducible representation of the HOMO.}
\label{fig_ex2}
\end{figure}

\clearpage
\newpage

\begin{figure}
\centering
\includegraphics[scale=0.75]{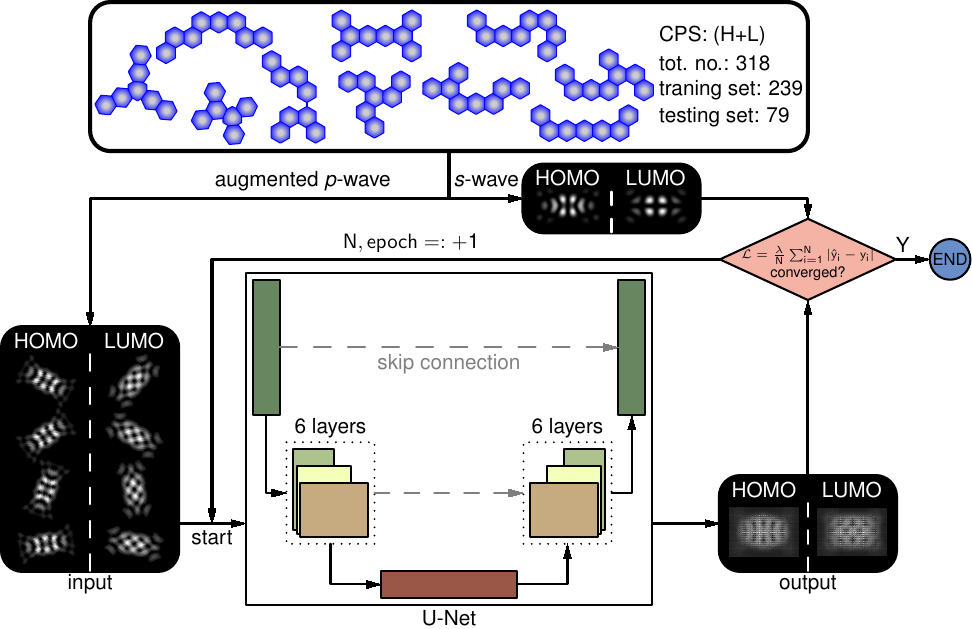}
\caption{\textbf{STM-Net architecture for a pure \textit{p}-wave tip}.
Schematic of the STM-Net training process for the transformation of STM
images under a pure \textit{p}-wave tip to the corresponding images
under an \textit{s}-wave tip (see details in Methods).}
\label{fig_ex3}
\end{figure}

\clearpage
\newpage

\begin{figure}
\centering
\includegraphics[scale=0.75]{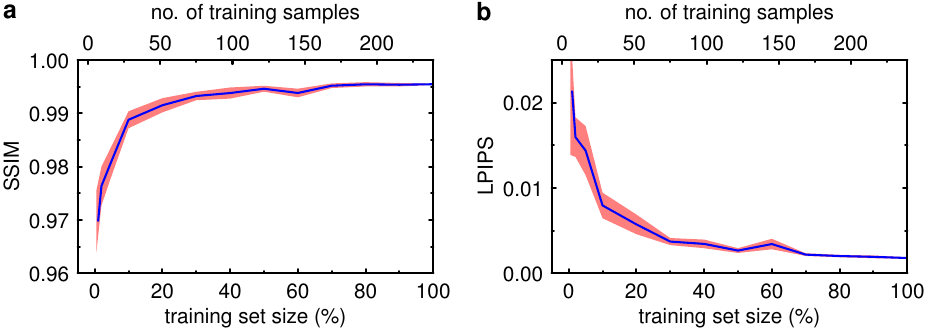}
\caption{\textbf{Other evaluation metrics for the \textit{p}-to-\textit{s}
STM-Net}. The learning curves for SSIM (\textbf{a}) and LPIPS
(\textbf{b}) of the \textit{p}-to-\textit{s} STM-Net. The pink-shaded
region represents the 95\% confidence interval obtained from 10 separate
training runs. The blue line shows the mean metrics values averaged
across these 10 training runs.}
\label{fig_ex4}
\end{figure}

\end{document}